\documentstyle[aps]{revtex}

\newcommand{\beq}{\begin{equation}}

\newcommand{\eeq}{\end{equation}}

\newcommand{\four} {  {}^{(4)}\kern-1pt  }

\newcommand{\ben}{\begin{eqnarray}}

\newcommand{\een}{\end{eqnarray}}

\catcode`@=11     
\def\eqalign#1{\null\,\vcenter{\openup\jot\m@th
  \ialign{\strut\hfil$\displaystyle{##}$&$\displaystyle{{}##}$\hfil
      \crcr#1\crcr}}\,}
\catcode`@=12   
\begin{document}
\title{\begin{flushright}
hep-th/0109191 \\
$~$ \\
\end{flushright}
PHENOMENOLOGY OF PARTICLE PRODUCTION AND PROPAGATION IN \\
STRING-MOTIVATED CANONICAL NONCOMMUTATIVE SPACETIME}
\author{{\bf Giovanni Amelino-Camelia} and {\bf Luisa Doplicher}}
\address{Dipart.~Fisica,
Univ.~Roma ``La Sapienza'', and INFN Sez.~Roma1\\
P.le Moro 2, 00185 Roma, Italy}
\author{{\bf Soonkeon Nam} and {\bf Yun-Seok Seo}}
\address{Dept.~of Physics and Basic Sciences Research Institute\\
Kyung Hee University, Seoul, 130-701, Korea} \maketitle
\begin{abstract}
We outline a phenomenological programme for the search of effects
induced by (string-motivated) canonical noncommutative spacetime.
The tests we propose are based, in analogy with a corresponding
programme developed over the last few years for the study of
Lie-algebra noncommutative spacetimes,
on the role of the noncommutativity parameters
in the $E(p)$ dispersion relation.
We focus on the role of deformed dispersion relations
in particle-production collision processes, where the noncommutativity
parameters would affect the threshold equation,
and in the dispersion of gamma rays observed
from distant astrophysical sources.
We emphasize that the studies here proposed
have the advantage of involving particles
of relatively high energies, and may therefore be
less sensitive to ``contamination" (through IR/UV mixing)
from the UV sector of the theory.
We also explore the possibility
that the relevant deformation of the dispersion
relations could be responsible for the experimentally-observed violations
of the GZK cutoff for cosmic rays and could have a role in the
observation of hard photons from distant astrophysical sources.
\end{abstract}

\section{Introduction and summary}
Most approaches to the unification of
general relativity and quantum mechanics lead to the emergence,
in one or another way, of noncommutative geometry.
While one can in principle consider a relatively
wide class~\cite{connesbook,starwess}
of flat noncommutative spacetimes (``quantum Minkowski"),
most studies focus on the two
simplest examples; ``canonical noncommutative spacetimes"
($\mu,\nu,\beta = 0,1,2,3$)
\begin{equation}
[x_\mu,x_\nu] = i \theta_{\mu,\nu}
\label{canodef}
\end{equation}
and ``Lie-algebra noncommutative spacetimes"
\begin{equation}
[x_\mu,x_\nu] = i C^\beta_{\mu,\nu} x_\beta ~.
\label{liedef}
\end{equation}
The canonical type (\ref{canodef}) was originally proposed~\cite{dopl}
in the context of attempts to develop a new fundamental picture
of spacetime\footnote{But in order to play a role in a fundamental
picture of spacetime the $\theta_{\mu,\nu}$
cannot be constants ({\it e.g.}
they should themselves be elements of an enlarged algebra, together
with the coordinates~\cite{dopl}).}.
More recently, (\ref{canodef}) is proving useful in the
description of string theory in presence of certain backgrounds
(see, {\it e.g.},
Refs.~\cite{Connes:1998cr,Douglas:1998fm,Chu:1999qz,Seiberg:1999vs,suss,Chu:2000ww,Douglas:2001ba}).
String theory in these backgrounds admits description (in the sense
of effective theories) in terms of a field theory in the noncommutative
spacetimes (\ref{canodef}), with the tensor $\theta_{\mu,\nu}$
reflecting the properties of the specific background.
Among Lie-algebra, type (\ref{liedef}), noncommutative versions
of flat (Minkowski) spacetime, research
has mostly focused on $\kappa$-Minkowski~\cite{majrue,kpoinap,gacmaj}
spacetime ($l,m = 1,2,3$)
\begin{equation}
[x_m,t] = i \lambda x_m ~,~~~~[x_m, x_l] = 0 ~.
\label{kmindef}
\end{equation}
This form of noncommutativity can provide, for fixed value of the scale
$\lambda$, a new fundamental description of (flat)
spacetime~\cite{dsr1dsr3}, which in particular hosts a maximum
momentum~\cite{dsr2,jurekdsr,gacrosjurek}. No effective-theory applications
({\it e.g.}, again for the description of physics in some background
fields) has yet been found for (\ref{kmindef}).

Several authors have
proposed techniques that allow to place experimental bounds on the
(dimensionful) parameters
$\theta_{\mu,\nu}$~\cite{Chaichian:2001si,rizzo,Carroll:2001ws,Guralnik:2001ax,dine,Carlson:2001sw} and
$\lambda$~\cite{dsr1dsr3,kpoinplb,polonpap1,ita,gactp,abovethre}.
The strategies followed on the two sides, limits on the $\theta_{\mu,\nu}$
of canonical spacetime and limits on the $\lambda$ of $\kappa$-Minkowski
Lie-algebra spacetime, have been rather different and this is partly
understandable in light of the differences between the two scenarios,
particularly
\begin{itemize}
\item The fact that (\ref{kmindef}) could provide a fundamental picture of
spacetime encourages the assumption that $\lambda$ be of the order of the
Planck length $L_p \simeq 1.6 {\cdot} 10^{-33} cm$, so the search of experimental
tests is naturally aiming for corresponding sensitivities. Instead, in its
popular string-theory application, the $\theta_{\mu,\nu}$ parameters of
(\ref{canodef}) reflect the properties of a background field of the
corresponding string-theory context, and therefore there is no natural
estimate for the $\theta_{\mu,\nu}$ (they will depend on the strength of
the background field).
\item The new effects predicted by field theories
in canonical spacetime (\ref{canodef}) could be observably large
also in the low-energy regime.
Instead field theories in the Lie-algebra spacetime (\ref{kmindef})
predict noncommutativity effects that
are vanishingly small for low-energy particles,
but become
significant in the high-energy regime (where the particle/probes have
wavelength short enough to be affected by the small, $\lambda$-suppressed,
noncommutative effects).
\item As one easily realizes based on the fact that the
$\theta_{\mu,\nu}$ of (\ref{canodef}) reflect the properties of a
background in the corresponding string-theory context, the
$\theta_{\mu,\nu}$ parameters cannot be treated as observer-independent
(the string-theory background takes different form/value in different
inertial frames and the $\theta_{\mu,\nu}$ transform accordingly). This
must be taken into account in the analysis of the experimental contexts
that could set bounds on the $\theta_{\mu,\nu}$, and in particular it has
important implications for the task of combining the limits obtained by
different experiments\footnote{Limits on the $\theta_{\mu,\nu}$ obtained
using data analyses that assume different inertial frames cannot be
directly combined/compared; one must first transform all analyses into a
single inertial frame ({\it e.g.} the one naturally identified by the
cosmic microwave background). This is unfortunately not taken into
account by some
authors.}. Instead, as clarified in Refs.~\cite{dsr1dsr3,dsr2},
noncommutativity of type (\ref{kmindef}) does not identify a preferred
inertial frame ($\lambda$ is observer-independent), but rather reflects a
deformed action of the boost generators. All limits on $\lambda$ can
therefore be combined straightforwardly.
\end{itemize}
In spite of the fact that indeed these differences between the two
noncommutativity scenarios are rather significant, in this paper we show
that some of the phenomenological contexts that are being analyzed
from a $\kappa$-Minkowski,  (\ref{kmindef}), perspective
can also deserve interest from
the perspective of canonical noncommutativity (\ref{canodef}).
The key ingredient for the experimental studies we consider
is a deformed dispersion relation. Deformed dispersion relations arise
naturally in noncommutative spacetimes. The conventional
special-relativistic
dispersion relation $E^2 = {\vec p}^2 + m^2$ reflects the classical
Lorentz symmetries of classical Minkowski spacetime. Noncommutative
versions of Minkowski spacetime do not enjoy the same classical
symmetries, and they therefore naturally lead to deformed dispersion
relations. The $\kappa$-Minkowski spacetime (\ref{kmindef}) is known to be
invariant~\cite{dsr1dsr3} under a group of deformed Lorentz
transformations; there is no loss of symmetries (infinitesimal Lorentz
transformations still correspond to 6 generators, which were already well
understood in the mathematical-physics studies of deformed Poincar\'{e}
algebras~\cite{kpoinap,kpoinold}), but the nature of these symmetries is
different from the classical case and there is a corresponding deformation
of the dispersion relation. Canonical noncommutative spacetimes
(\ref{canodef}) are clearly less symmetric than classical Minkowski
spacetime (again, this
 is easily understood considering the presence of a background field in the
corresponding string-theory picture). The loss of symmetries is
reflected in a deformation of the dispersion relation\footnote{Besides the
fact that in $\kappa$-Minkowski Lie-algebra spacetime there is no symmetry
loss (only symmetry deformation), while in canonical noncommutative
spacetimes there is a net loss of symmetries, another important difference
is the fact that in field theories on $\kappa$-Minkowski the deformation of
the dispersion relation is already encountered at tree level, while field
theories on canonical noncommutative spacetimes have
an unmodified ({\it i.e.} special-relativistic) tree-level propagator
and the dispersion-relation deformation emerges only
through loop corrections.}.

Our proposal of investigating the implications of
the deformed dispersion relations of
the canonical case (\ref{canodef}) in the same high-energy contexts
previously considered for the
deformed dispersion relations of $\kappa$-Minkowski might
at first appear surprising.
Whereas the significance of the
dispersion-relation deformations encountered in $\kappa$-Minkowski
increases with energy, the most significant dispersion-relation-deformation
effects in canonical noncommutative spacetimes are
formally found at low energies.
But, as we show here, the relevant observational constraints are
significant for canonical-noncommutativity in spite of the
relatively high energies involved.
We therefore argue that
these observations may be relevant for several scenarios
based on canonical noncommutativity.
In particular, these observations appear to be relevant
in attempts to test the idea of canonical noncommutativity in regimes
which are immune from the ``contamination" of unknown features
of the UV sector. If unknown new structures come in at a UV scale $\Lambda$
they can also affect (through IR/UV mixing)
the predictions of the theory at low-energy scales.

We adopt in this paper
a strictly phenomenological viewpoint, but we are concerned with
the rather preliminary status of the understanding of the
infrared problems of field theories in canonical noncommutative
spacetimes. Limits on $\theta_{\mu,\nu}$ are being set in some
studies based on low-energy considerations, but the IR/UV mixing
introduces in these low-energy predictions a strong sensitivity
(see, {\it e.g.}, Refs.~\cite{dine,gianlucaken})
to the structure of the unknown UV sector.

Some of the anomalous effects encountered in naive low-energy
studies could be made ``even worse" (more anomalous with respect
to conventional theories) as  a result of effects induced
through the IR/UV mixing~\cite{dine}.
It is also conceivable that the opposite might happen:
the unknown UV sector might have such special properties
that it ``cures" (through the IR/UV mixing) the
anomalous features encountered in naive low-energy analysis.
For example, it is sufficient to have SUSY (supersymmetry)
in the UV sector to remove~\cite{suss} several
troblesome low-energy predictions of canonical noncommutativity,
and on the basis of the SUSY example it appears
plausible~\cite{gianlucaken} that
other features of the UV sector might positively affect the
nature of the low-energy predictions.

High-energy data are already
important, in theories in commutative spacetime,
as a way to test theories in different regimes,
providing information complementary to that obtained in
low-energy studies, but in the case of canonical noncommutative
spacetimes high-energy observations might be even more significant,
since low-energy data provide indications which are subject to
possible modification as a result of IR/UV mixing.
In order to give a more explicit description of the type of
scenarios we are considering,
let us consider an example in which
\begin{description}
\item[(i)] The canonical-noncommutativity
scale\footnote{Because of the nature of the study we are
reporting, we find useful sometimes to consider a single
length scale $\sqrt{\theta}$ to characterize the matrix $\theta_{\mu,\nu}$.
Of course, in principle there can be large differences
between the non-zero entries of the $\theta_{\mu,\nu}$
matrix, and the line of argument we adopt is applicable also to
this more general situation (in some cases it would be sufficient
to take $\sqrt{\theta}$ as the largest length scale in $\theta_{\mu,\nu}$).}
$\sqrt{\theta}$
is of order $\sqrt{\theta} \sim 10^{-25}cm \sim (10^{11} GeV)^{-1}$.
\item[(ii)] SUSY (supersymmetry) is eventually ``restored'', at least
at some UV scale ${\cal M}_{susy}$
\item[(iii)]
The description of spacetime based exclusively on canonical
noncommutativity is only applicable up to
a UV cutoff $\Lambda \sim 10^{19} GeV$. For energies above the UV cutoff
spacetime acquires additional quantum features, even beyond
noncommutativity, that are perhaps described by string theory or
some other candidate ``quantum gravity'' theory. At the scale
$\Lambda$ we could also expect additional particle degrees of
freedom.
\end{description}
In such a scenario the low-energy
regime (say, $E<10^3 GeV \sim 1/(\Lambda \theta)$)
could be largely unaffected by the
spacetime noncommutativity, if the properties of the UV sector
(through the IR/UV mixing) conspired to largely cancel out the
characteristic low-energy predictions of canonical noncommutativity
(in the same way and beyond what is understood to be the effect of
UV SUSY on the low-energy sector of canonical noncommutative theories).
Since the UV ($E > \Lambda$) sector is unknown and its features
affect low-energy phenomena it is indeed plausible that the net
result of noncommutativity combined with new UV physics would be
giving us back physical predictions that are very close to the
ones of an ordinary low-energy theory in commutative spacetime.
This may require\footnote{Actually, the
necessary fine-tuning of the UV sector might not even
have to be very large~\cite{gianlucaken}:
as mentioned,
simply assuming that SUSY is indeed restored at
some UV scale already leads to
significant reduction of the ``anomalies" that would otherwise
affect theories in canonical noncommutative spacetime in the
low-energy regime.} large fine-tuning of the UV sector, but from a
strictly phenomenological perspective it cannot be excluded,
and it cannot be excluded if one is aiming for unconditional
high-confidence bounds on the noncommutativity parameters.

So in the scenario characterized by
hypotheses {\bf (i)},{\bf (ii)},{\bf (iii)}
the theory in canonical noncommutative
spacetime might be completely indistinguishable from a
corresponding theory in commutative spacetime at low-energy
scales, scales such that $E < 1/(\theta \Lambda)$.
But above the scale $1/(\theta \Lambda)$ the theory
in canonical noncommutative
spacetime is ``protected" from the UV: the predictions of a
theory in canonical noncommutative spacetime
in the energy range $1/(\theta \Lambda) < E < 1/(\sqrt{\theta})$
are largely insensitive to the UV sector.
We are not claiming that
this scenario in which low-energy predictions are unaffected
(as a result of mixing with a fine-tuned UV sector)
by canonical noncommutativity is the only possibility
(perhaps indeed it only occurs upon severe fine-tuning of
the UV sector), but we are interested here in this particular
possibility, especially since it would represent a
huge challenge from a phenomenological perspective.

In the next Section we discuss (relying in part on results which had already
appeared in the literature) the emergence of deformed dispersion relations
in field theories (mostly QED) on canonical noncommutative spacetimes. We
emphasize that this effect is particularly significant for uncharged
particles; in fact, in canonical noncommutative spacetimes it is possible
to describe the particles that we observe as neutral with respect to the
electromagnetic interactions as
particles that do interact with the photon (as a result of the structure of
the ``star product''~\cite{suss}) in contexts in which
the $\theta_{\mu,\nu}$ parameters cannot be neglected.
Loop corrections, and in particular photon dressing, of
neutral-particle propagators in canonical noncommutative spacetimes
generally induce corrections to the dispersion relation.

In Section~III, in preparation for the later discussion of
experimental tests, we discuss the type of contexts in which our
phenomenological strategy appears to have greater potential for
insight.

In Section~IV
we outline a phenomenological programme which can
lead to  bounds on (or estimates of)
the $\theta_{\mu,\nu}$ parameters by studying the
implications of the deformed dispersion relations discussed in Section~II.
We focus on the implications of a
deformed dispersion relation $E^2 = f(p) \neq {\vec p}^2 + m^2$ for:
(E1) the threshold condition for particle production
in certain collision processes, and (E2)
the dispersion of signals observed
from distant astrophysical sources. Since our objective here is the one of
outlining a relatively wide phenomenological programme, postponing to
future studies a detailed analysis of each of the proposals, we just list a
few experimental contexts and provide estimates of the sensitivity
levels that appear to be within reach of these experimental contexts.
Concerning the implications of the
deformed dispersion relations for the dispersion of signals observed from
distant astrophysical sources, we focus on the case of gamma-ray
bursts~\cite{piranGRBs}, whose observation provides powerful constraints on
dispersion for photons, and on the case of 1987a-type supernovae, which can
be used to constrain possible deformations of the dispersion relation for
neutrinos. Concerning threshold conditions,
we argue that high sensitivity to
nonvanishing values of the $\theta_{\mu,\nu}$
parameters can be achieved by analyzing
the threshold for electron-positron pair production from photon-photon
collisions at energy scales relevant for the expected cutoff on the
energies of hard photons observed from distant astrophysical sources.
Similarly, deformed thresholds associated with dispersion-relation
deformations can significantly affect the GZK cutoff for the
observation of cosmic rays. Ultra-high-energy cosmic-ray protons,
with energies in excess of the GZK cutoff, should not be detected
by our observatories because they should lose energy (thereby
complying with the GZK cutoff) through photo-pion production off
cosmic-microwave-background photons. If the present (commutative
spacetime) estimates of the GZK hard-proton cutoff and of the
analogous hard-photon cutoff were confirmed experimentally, we
could obtain very stringent limits on the $\theta_{\mu,\nu}$.
Interestingly, observations of cosmic rays presently
appear~\cite{kifu,gactp} to be in conflict with the cutoff
estimates that assume the conventional dispersion relation (and
the associated classical picture of spacetime).

Finally, Section~V is devoted to some closing remarks,
particularly concerning the outlook of the phenomenological programme here
outlined.

\section{Deformed dispersion relations in canonical noncommutative spacetime}
The subject of field theory in canonical noncommutative spacetimes
has been extensively studied, and there is a wide literature where
the reader can find rather pedagogical
introductions (see, {\it e.g.},
Refs.~\cite{suss,Filk:1996dm,Chaichian:2000kp,Sheikh-Jabbari:1999iw,Bigatti:2000iz,Ishibashi:2000hs,Minwalla:2000px,Hayakawa:2000yt,Aref'eva:2001bg}).
We here just mention some well-established features of these
field theories, and focus
on a few loop corrections that are relevant for the analysis
of the emergence of deformed dispersion relations.\footnote{The loop
corrections we are interested in concern the photon self-energy
at one loop, and the one-loop contributions involving the photon
to the self-energies of ``neutral" spin-1/2 fermions
and ``neutral" spin-0 bosons. Our results on the photon self-energy
reproduce the ones of the previous analysis reported in Ref.~\cite{suss}.
We did not find in the literature any analogous analysis of
the self-energy of neutral spin-1/2 fermions.
For the self-energy of neutral spin-0 bosons
a related analysis has been reported in Ref.~\cite{Aref'eva:2001bg};
however, that study adopted a different gauge choice and it is
not clear to us (it was not explicitly stated) whether it
concerned ``neutral" or ``charged" spin-0 bosons (if it was meant to
consider neutral particles, it would investigate the same loop corrections
which are here of interest and our results would be in disagreement with some
of the results of Ref.~\cite{Aref'eva:2001bg},
even taking into account the different choice of gauge).}
The tree-level propagators are unmodified by the noncommutativity.
Loop corrections to the two-point functions introduce terms
that correspond to deformations of the dispersion relations and reflect
the loss of symmetry associated with nonvanishing $\theta_{\mu,\nu}$
parameters. Particularly rich are the structures of the dressed photon
propagator~\cite{suss} and of the photon corrections to the propagators
of particles that are neutral in the commutative-spacetime limit.
Certain gauge-invariant actions in
canonical noncommutative spacetime describe particles that are coupled
to the gauge field only for nonvanishing $\theta_{\mu,\nu}$,
{\it i.e.} these
are particles that in the $\theta_{\mu,\nu} \rightarrow 0$ limit
no longer interact with the gauge field. These particles are natural
candidates to describe the neutral particles that we observe.
In the remainder of this section we discuss the $\theta_{\mu,\nu}$
deformation of the dispersion relations for photons, neutral
spin-1/2 fermions, and neutral spin-0 bosons. Here and in the
following we are describing as (QED-)``neutral" the type of
particles described above (no interactions with the photon in the
$\theta_{\mu,\nu} \rightarrow 0$ limit, but some
$\theta_{\mu,\nu}$-dependent interactions with the photon in the
noncommutative spacetime). We work in Feynman gauge, we adopt the
standard notation \beq\label{ptild} {\tilde p}_\mu \equiv
p^\alpha \theta_{\mu,\alpha}~, \eeq and we also assume throughout
that $\theta_{\mu,0} =0= \theta_{0,\mu}$, {\it i.e.} we consider
the case in which only (some of) the space components
$\theta_{i,j}$ are nonvanishing\footnote{The case of space/time
noncommutativity ($\theta _{0i}\neq 0$) is not necessarily void
of interest~\cite{stgood}, but it is more delicate, especially in
light of possible concerns for unitarity~\cite{unitarity}. Since
our analysis is not focusing on this point we will simply assume
that $\theta _{0i}=0$.}. Although we are planning (as discussed
in Section~I) to develop a phenomenology for field theory in
canonical noncommutative spacetime at certain relatively high
energies, we will not explicitly focus on these energy scales in
this section. However, as discussed in Sections~III and IV, our
phenomenological programme is primarily sensitive to the leading
$\theta$-dependent deformations of the dispersion relations, and
the fact that we intend to analyze contexts involving relatively
high energy scales allows us to analyze self-energies with the
implicit assumption that terms of order $\Lambda^{-2}$ can be
neglected with respect to terms of order $p^2 \theta^2$, when $p$
is an external momentum ($p > 1/(\theta \Lambda)$). To the cutoff
scale $\Lambda$ we attribute consistently the interpretation
discussed in Section~I: we assume that at some scale $\Lambda$
our field theories in canonical noncommutative spacetime fail to
apply in a rather significant way, perhaps as a result of the
presence of new stringy or quantum-gravity effects, and that,
because of the IR/UV mixing~\cite{suss,gianlucaken},
this unknown UV physics could affect
profoundly the predictions of the theory at scales below
$1/(\theta \Lambda)$.

\subsection{Deformed dispersion relation for photons}
In order to see the emergence of a deformed dispersion relation for photons
in canonical noncommutative spacetimes it is sufficient to consider
one-loop contributions to the photon propagator from diagrams involving as
virtual particles either photons themselves or other ``neutral" particles.
We shall often use the notation $\gamma$ for photons, $\nu$ for neutral
spin-1/2 fermions, and $\Phi$ for neutral spin-0 bosons. Some relevant
interaction vertices of the field theory in canonical noncommutative
spacetime are:
\begin{itemize}
\item the four-$\gamma$  vertex
\ben
-4ig^{2}&&\left( \left(
g^{\alpha \gamma}g^{\beta \delta}-g^{\alpha
\delta}g^{\beta \gamma}\right)  \sin\frac{\tilde{p}_{1}p_{2}}{2}\sin
\frac{\tilde{p}_{3}p_{4}}{2}+ \right. \nonumber \\
&&+\left(
g^{\alpha \delta}g^{\beta \gamma}-g^{\alpha \beta} g^{\gamma
\delta}\right)
\sin\frac{\tilde{p}_{3}p_{1}}{2}\sin\frac{\tilde{p}_{2}p_{4}}{2}+
\label{vgggg} \\
&&+\left.\left( g^{\alpha \beta}
g^{\gamma \delta}-g^{\alpha \gamma}g^{\beta \delta}\right)
\sin\frac{\tilde{p}_{1}p_{4}}{2}\sin\frac{\tilde{p}_{2}p_{3}}{2}\right)
\nonumber
\een
with every $p_{i}$ exiting the vertex;
\item the three-$\gamma$  vertex
\beq\label{vggg}
-2g\sin\frac{\tilde{p}_{1}p_{2}}{2}\left(g^{\alpha \beta}\left(
p_{1}-p_{2}\right)^{\gamma}+g^{\alpha \gamma}\left(p_{3}-
p_{1}\right)^{\beta}+g^{\beta \gamma}\left(p_{2}-p_{3}\right)^{\alpha}\right)
\eeq
with every $p_{i}$ exiting the vertex;
\item the $\gamma$-$\nu$-$\nu$ vertex
\beq\label{vgnn}
2g\gamma^{\mu}\sin\frac{\tilde{p}_{1}p_{2}}{2}
\eeq
where $p_{1}$ ($p_{2}$) is the momentum of the incoming (outgoing)
$\nu$;
\item the $\gamma$-$\gamma$-$\Phi$-$\Phi$ vertex
\beq\label{vggff}
4ig^{2}g^{\mu\nu}\left(\sin\frac{\tilde{p}_{2}p_{4}}{2}\sin
\frac{\tilde{p}_{1}p_{3}}{2}+\sin\frac{\tilde{p}_{2}p_{3}}{2}\sin\frac
{\tilde{p}_{1}p_{4}}{2}\right)
\eeq
where $p_{1}$, $p_{2}$ are the photons' momenta and $p_{3}$, $p_{4}$ the
scalars' momenta, and they all exit the vertex;
\item and the $\gamma$-$\Phi$-$\Phi$ vertex
\beq\label{vgff}
2 g \left(p_{1}+p_{2}\right)^{\mu}\sin\frac{\tilde{p}_{1}p_{2}}{2}
\eeq
where $p_{1}$ ($p_{2}$) is the momentum of the incoming (outgoing)
scalar.
\item In addition we will also need the $\gamma$-ghost-ghost
vertex\footnote{It is well known that abelian gauge theories
in canonical noncommutative
spacetime, unlike their commutative counterparts, do require the
introduction of ghosts. This and other properties render abelian gauge
theories in canonical noncommutative spacetime somewhat analogous to
nonabelian gauge theories in commutative spacetime.}:
\beq\label{vghh}
2gp_{2}^{\mu}\sin\frac{\tilde{p}_{1}p_{2}}{2}
\eeq
where $p_{1}$ ($p_{2}$) is the momentum of the incoming (outgoing) ghost.
\end{itemize}

There are three
nontrivial pure-gauge one-loop contributions to the photon self-energy. For
external photons with momentum $p$ and polarizations $\mu$ and $\nu$ one
finds
\begin{itemize}
\item a tadpole-type diagram, using the vertex (\ref{vgggg}),
\beq\label{sgvgggg}
i \Sigma_{\gamma;\gamma;1}^{\mu\nu}
 =- 2 ig^{2}\int\frac{d^{4}k}{\left(2\pi\right)^{4}}
\frac{-ig_{\rho\sigma}}{k^{2}}\left(2g^{\mu\nu
}g^{\rho\sigma}-g^{\mu\sigma}g^{\nu\rho}
-g^{\mu\rho}g^{\nu\sigma}\right)\sin^{2}\frac{\tilde{p}k}{2},
\eeq
\item a photon-loop diagram, using the vertex (\ref{vggg}),
\ben
i \Sigma_{\gamma;\gamma;2}^{\mu\nu}
&=& 2 g^{2}\int\frac{d^{4}k}{\left(2\pi \right) ^{4}}
\frac{-ig_{\rho \alpha }}{k^{2}}\left( g^{\mu \sigma
}\left( p-k\right) ^{\rho }+g^{\mu \rho }\left( -2p-k\right) ^{\sigma
}+g^{\rho \sigma }\left( 2k+p\right) ^{\mu }\right)    \nonumber \\
&& \frac{-ig_{\sigma \beta }}{\left( p-k\right) ^{2}}\left( g^{\nu
\beta }\left( -p+k\right) ^{\alpha }+g^{\nu \alpha }\left( 2p+k\right)
^{\beta }+g^{\alpha \beta }\left( -p-2k\right) ^{\nu }\right)\sin ^{2}
\frac{\tilde{p}k}{2}, \label{sgvggg}
\een
\item and a ghost-loop diagram, using the vertex (\ref{vghh}),
\beq\label{sgvghh}
i \Sigma_{\gamma;ghost}^{\mu\nu}
 =4g^{2}\int\frac{d^{4}k}{\left( 2\pi\right) ^{4}}
\frac{i}{k^{2}}\left( p+k\right) ^{\mu}\frac{i}{\left( p-k\right) ^{2}}
k^{\nu}\sin^{2}\frac{\tilde{p}k}{2}.
\eeq
\end{itemize}
With straightforward calculations one finds~\cite{suss} that the
dominant $\theta$-dependent\footnote{Here and in the following we
describe generically as ``dominant $\theta$-dependent
contributions" the contributions that are most significant at
scales that are greater than $1/(\theta \Lambda)$
(so that $1/\Lambda^2 \ll (p \theta)^2$) but are not much greater
than $1/(\theta \Lambda)$ (so that terms of order $1/(p \theta)^2$
are not completely negligible in comparison with $p^2$).
In practice this means that we aim at
a description of scales greater than $1/(\theta \Lambda)$, but we
still intend to focus on contributions that come from the part of
the loop integration that involves large loop momenta.}
contribution from these diagrams are
\beq\label{sgvggggres}
i \Sigma_{\gamma;\gamma;1}^{\mu\nu}
 =\frac{3ig^{2}g^{\mu\nu}}{
2\pi^{2}\left| \tilde {p}\right| ^{2}},
\eeq
\beq\label{sgvgggres}
i \Sigma_{\gamma;\gamma;2}^{\mu\nu}
=\frac{-ig^{2}}{4\pi^{2}}\left(
7\frac{g^{\mu\nu}}{\left| \tilde
{p}\right| ^{2}}-10\frac{\tilde{p}^{\mu}\tilde{p}^{\nu}}{\left| \tilde
{p}\right| ^{4}}\right),
\eeq
and
\beq\label{sgvghhres}
i \Sigma_{\gamma;ghost}^{\mu\nu}
 =\frac{ig^{2}}{4\pi^{2}}\left(
-2\frac{\tilde{p}^{\mu}\tilde{p}^{\nu}}{%
\left| \tilde{p}\right| ^{4}}+\frac{g^{\mu\nu}}{\left| \tilde{p}\right|
^{2}}%
\right).
\eeq

Next let us consider the one-loop contribution to the photon self-energy
that involves virtual neutral fermions, using the vertex (\ref{vgnn}):
\beq\label{sgvgnn}
i \Sigma_{\gamma;\nu}^{\mu\nu}
 =4g^{2}\int\frac{d^{4}k}{\left( 2\pi\right)
^{4}}\sin^{2}\frac{\tilde{p}k}{2}%
tr\left( \gamma^{\mu}\frac{i}{\gamma^{\rho}\left( p+k\right) _{\rho }-m}%
\gamma^{\nu}\frac{i}{\gamma^{\sigma}k_{\sigma}-m}\right).
\eeq
from which we extract again the dominant $\theta$-dependent contribution
\beq\label{sgvgnnres}
i \Sigma_{\gamma;\nu}^{\mu\nu}
=-\frac{4ig^{2}}{\pi^{2}}\frac{\tilde{p}^{\mu}\tilde{p}^{\nu}}{\left|
\tilde{p}\right| ^{4}}.
\eeq
Finally we consider the two nontrivial one-loop contributions to the photon
self-energy that involve virtual neutral scalars: a tadpole-type diagram,
using the vertex (\ref{vggff}),
\beq\label{sgvggff}
i \Sigma_{\gamma;\Phi;1}^{\mu\nu}
 =4ig^{2}g^{\mu\nu}\int\frac{d^{4}k}{\left(
2\pi\right) ^{4}}\frac{i}{
k^{2}-m^{2}}\sin^{2}\frac{\tilde{p}k}{2},
\eeq
and a scalar-loop diagram, using the vertex (\ref{vgff}),
\beq\label{sgvgff}
i \Sigma_{\gamma;\Phi;2}^{\mu\nu}
 =-2g^{2}\int\frac{d^{4}k}{\left( 2\pi\right)
^{4}}\frac{i}{k^{2}-m^{2}}\frac{
i}{\left( p+k\right) ^{2}-m^{2}}\left( p+2k\right) ^{\mu}\left(
p+2k\right)
^{\nu}\sin^{2}\frac{\tilde{p}k}{2},
\eeq
whose dominant $\theta$-dependent contributions are
\beq\label{sgvggffres}
i \Sigma_{\gamma;\Phi;1}^{\mu\nu}
 =\frac{ig^{2}g^{\mu\nu}}{2\pi^{2}\tilde{p}^{2}},
\eeq
and
\beq\label{sgvgffres}
i \Sigma_{\gamma;\Phi;2}^{\mu\nu}
 =-\frac{ig^{2}}{2\pi^{2}}\left(
-2\frac{\tilde{p}^{\mu}\tilde{p}^{\nu}}{ \left| \tilde{p}\right|
^{4}}+\frac{g^{\mu\nu}}{\left| \tilde{p}\right| ^{2}} \right).
\eeq Combining these results one obtains~\cite{suss} the total
contribution to the photon self-energy: \beq\label{suss310} i
\Sigma^{\mu\nu}_\gamma ={ig^{2} \over \pi^{2}} \left(
N_{s}+2-2N_{f}\right)
\frac{\tilde{p}^{\mu}\tilde{p}^{\nu}}{\tilde{p}^{4}}. \eeq where
$N_s$ and $N_f$ denote the number of neutral scalar fields and the
number of neutral fermion fields in the theory. It is important to
notice~\cite{suss} that in a supersymmetric field theory in
canonical noncommutative spacetime, which would accommodate an
equal number of bosonic and fermionic degrees of freedom, this
correction term vanishes. However, while we do want to emphasize
the implications that apply in particular to supersymmetric
theories, we are here interested in general in the predictions of
field theories in canonical noncommutative spacetime. There is at
present no direct experimental evidence of supersymmetry, and
accordingly in our phenomenological analysis we will assume that
either there is no supersymmetry at all or that supersymmetry is
broken at low energies (processes below the
supersymmetry-restoration scale ${\cal M}_{susy}$, here treated
as a phenomenological parameter). It is therefore important from
our perspective to explore the physical content of
(\ref{suss310}) when supersymmetry is absent (or not yet
restored). It is convenient to consider the simple case in which
$\theta_{i,j}$ is only nontrivial in the (1,2)-plane:
$\theta_{1,2} = - \theta_{2,1} \equiv \theta \neq 0$,
$\theta_{1,3} = \theta_{2,3} =0$. Relevant observations have
already been reported in Ref.~\cite{suss}: according to
(\ref{suss310}) the two, transversely polarized, physical degrees
of freedom of the photon satisfy different dispersion relations,
reflecting the loss\footnote{On this issue of the breaking of
Lorentz invariance there is sometimes some confusion. At the
fundamental level these theories in canonical noncommutative
spacetimes are still special-relativistic in the ordinary sense.
However, $\theta$ has the role of a background (as well
understood in the corresponding string-theory picture) and of
course effective field theories in presence of some background do
not enjoy Lorentz symmetry. The Lorentz invariance of the theory
becomes manifest only in formalisms that take into account both
the transformations of the quantum fields and of the background.
The effective field theory does have a preferred frame (which one
can identify by selecting, {\it e.g.}, a frame in which the form
of the background is particularly simple) but this is of course
fully consistent with special relativity. The situation is
completely different in Lie-algebra noncommutative spacetimes.
For example, it was recently shown~\cite{dsr1dsr3,dsr2} that
$\kappa$-Minkowski does not reflect the properties of a
background and does not have a preferred frame, both $\lambda$
and $c$ have in $\kappa$-Minkowski roles which are completely
analogous to the role that only $c$ enjoys in classical
Minkowski: the role of an observer-independent kinematic
scale~\cite{dsr1dsr3,dsr2}, left invariant under (deformed)
Lorentz transformations (and in particular the
observer-independent scale $\lambda$ acquires the role of the
inverse of the maximum momentum of the
theory~\cite{dsr2,jurekdsr,gacrosjurek}). Instead in canonical
noncommutative spacetimes the $\theta_{\mu,\nu}$ is not invariant
under Lorentz transformations; it transforms of course like a
tensor. More on this type of considerations can be found in
Ref.~\cite{dsr1dsr3}.} of Lorentz invariance introduced by
$\theta$. If, for example, ${\tilde p}$ is in the 1-direction, in
the case we are considering, $\theta_{1,3} = \theta_{2,3} =
\theta_{\mu,0} =0$, one finds that the degree of freedom
polarized in the direction orthogonal to ${\tilde p}$ satisfies
the ordinary special-relativistic dispersion relation
\beq\label{dispgord} p_0^2 = {\vec p}^2 ~, \eeq while the degree
of freedom polarized in the direction parallel to ${\tilde p}$
satisfies a deformed dispersion relation of the type
\beq\label{dispgdef} p_0^2 = {\vec p}^2 + {\zeta_\gamma \over
{\tilde p}^2} ~, \eeq where $\zeta_\gamma$ is a number that
depends, according to (\ref{suss310}), on the coupling constant
and on the number of bosonic and fermionic degrees of freedom
present in the theory. Besides the polarization dependence, these
dispersion relations are strongly characterized by the term
$\zeta_\gamma /{\tilde p}^2$, which reflects the mentioned IR/UV
mixing.

\subsection{Deformed dispersion relation for neutral spin-0 bosons}
We now analyze in the same way the self-energy of scalars
(neutral spin-0 bosons) in canonical noncommutative spacetime.
We focus on the loop contributions involving virtual photons
and on the loop contributions involving virtual
scalars (self-interactions).
There are two nontrivial one-loop contributions to the scalar
self-energy that involve virtual photons (for a scalar of external
momentum $p$):
\begin{itemize}
\item a tadpole-type diagram, using the vertex (\ref{vggff}),
\beq\label{sfvggff}
i \Sigma_{\Phi;\gamma;1} =4ig^{2}g^{\mu\nu}\int\frac{d^{4}k}{\left(
2\pi\right) ^{4}}\frac{
-ig_{\mu\nu}}{k^{2}}\sin^{2}\frac{\tilde{p}k}{2},
\eeq
\item and a sunset-type diagram, using the vertex (\ref{vgff}),
\beq\label{sfvgff}
i \Sigma_{\Phi;\gamma;2}
=-4g^{2}\int\frac{d^{4}k}{\left( 2\pi\right) ^{4}}\left( p+k\right) ^{\mu
}
\frac{-ig_{\mu\nu}}{\left( p-k\right) ^{2}}\left( p+k\right)
^{\nu}\frac{i}{
k^{2}-m^{2}}\sin^{2}\frac{\tilde{p}k}{2}.
\eeq
\end{itemize}
Again, the dominant $\theta$-dependent contribution\footnote{In the
contexts analyzed in Section~IV one finds that terms of
the type $1/{\tilde p}^2$ are more significant than the ones of
the type $\ln({\tilde p}^2)$. In light of this observation we
consistently neglect logarithmic $\theta$ dependence when
terms of the type $1/{\tilde p}^2$ are present.}
from these diagrams
can be established with straightforward calculations,
finding
\beq\label{sfvggffres}
i \Sigma_{\Phi;\gamma;1} =-\frac{2ig^{2}}{\pi^{2}\left| \tilde{p}\right|
^{2}},
\eeq
and
\beq\label{sfvgffres}
i \Sigma_{\Phi;\gamma;2} =\frac{ig^{2}}{2\pi^{2}\left| \tilde{p}\right|
^{2}}.
\eeq
The one-loop (tadpole) contribution to the scalar self-energy due to
its self-interactions is a well-known prototype
result~\cite{suss,Minwalla:2000px} of field theory in canonical
noncommutative spacetime. One finds that the most important effect of the
noncommutativity in this self-interaction tadpole contribution is again of
the type $1/{\tilde p}^2$. Therefore combining photon dressing and self
dressing of the propagator one finds a deformed dispersion relation for
neutral spin-0 bosons of the type
\beq\label{dispgdefscalar}
p_0^2 = {\vec p}^2 + m_\Phi^2 + {\zeta_\Phi \over {\tilde p}^2}
~,
\eeq
where again $\zeta_\Phi$ is a number that depends on the coupling
constants of the theory and on the number and type of fields involved.

\subsection{Deformed dispersion relation for neutral spin-1/2 fermions}
We close this Section by performing an analogous dispersion-relation
analysis for neutral spin-1/2 fermions.
We only consider the self-energy contribution that is due to interactions
with the photon. The relevant sunset-type diagram,
using the vertex (\ref{vgnn}), corresponds to the integral
\beq\label{snvgnn}
i \Sigma_{\nu;\gamma} =-4g^{2}\int\frac{d^{4}k}{\left( 2\pi\right)
^{4}}\gamma^{\mu}\frac{i%
}{\gamma^{\rho}k_{\rho}-m}\gamma^{\nu}\frac{-ig_{\mu\nu}}{\left(
p-k\right) ^{2}}\sin^{2}\frac{\tilde{p}k}{2}; \eeq which can be
evaluated exactly, obtaining: \beq\label{iruveq} i
\Sigma_{\nu;\gamma} =\frac{2g^{2}}{\left( 4\pi\right) ^{2}}\left(
-\gamma^{\mu}p_{\mu }+4m\right)
\ln\frac{\Lambda^{2}}{\Lambda_{eff}^{2}}, \eeq where $\Lambda$ is
a conventional high-momentum cutoff and\footnote{The role of
$\Lambda_{eff}$ in Eq.~(\ref{iruveq}) and in other characteristic
formulas of field theory in canonical noncommutative spacetime,
obtained in previous studies, reflects the IR/UV mixing present
in these theories. If one takes $\Lambda \rightarrow \infty$ then
$\Lambda_{eff}^2 \rightarrow 4/\tilde{p}^{2}$ and $\ln
\Lambda_{eff}$ is singular in the infrared.} \beq\label{iruv}
\frac{1}{\Lambda_{eff}^{2}}=\frac{1}{\Lambda
^{2}}+\frac{\tilde{p}^{2}}{4}. \eeq The physical implications are
analogous to the ones found for photons and neutral spin-0 bosons.
For neutral spin-1/2 fermions one has again a deformation of the
dispersion relation which (if the fermion has mass) is singular
in the infrared; however, the singularity is softer, only
logarithmic, and accordingly the magnitude of the effects to be
expected (whether or not the fermion has mass) at relatively
small momenta is not as significant as, {\it e.g.}, for the
photon. We will therefore not express high expectations for the
bounds on $\theta$ that can be placed using observations of
neutral spin-1/2 fermions, but still we will comment on some
types of experiments/observations that are sensitive to the
dispersion-relation deformation experienced by these particles in
canonical noncommutative spacetime.

\section{Objectives of high-energy $\theta_{\mu,\nu}$ phenomenology}
In the next Section we discuss certain classes of observations in
astrophysics in which the deformed dispersion relations that are
characteristic of field theory in canonical noncommutative
spacetime can have significant implications.
These observations in astrophysics is involve particles
of relatively high energy (all relevant processes involve at
least one particle with $E > 1TeV$)
and the nature of the observations is such that they can be used
to constrain (or search for) even rather small deviations
from the standard $E^2 = m^2 + p^2$ dispersion relation.

We are bringing these observations to the attention of the community
interest in canonical noncommutativity since deformed dispersion relations
are a characteristic feature of field theories in
canonical-noncommutative spacetime.
These observations may therefore be relevant for several scenarios
based on canonical noncommutativity.
For example, as mentioned in Section~I,
these observations appear to be relevant
in attempts to test the idea of canonical noncommutativity in regimes
which are immune from the ``contamination" of unknown features
of the UV sector. If unknown new structures come in at a UV scale $\Lambda$
they can also affect (through IR/UV mixing)
the predictions of the theory at
energy scales below $1/(\theta \Lambda)$.
By looking for ultra-high-energy tests of canonical noncommutativity
one is essentially hoping to gain access to energy regimes that
are high enough to be  above the scale  $1/(\theta \Lambda)$.

As mentioned in Section~I, it is even plausible to contemplate
canonical-noncommutativity theories in which,
as a result of certain corresponding
hierarchies between the relevant energy scales and of
corresponding properties of the UV sector,
noncommutativity leads to negligible effects below the scale
$1/(\theta \Lambda)$, but
leads to the characteristic dispersion-relation-deformation
effects above the scale  $1/(\theta \Lambda)$.
Because of  the IR/UV mixing it cannot be excluded that
some properties of the (unknown) UV sector might lead to exact
(or nearly-exact) cancellations of the anomalous effects that
canonical noncommutativity would otherwise predict for processes
below the scale $1/(\theta \Lambda)$.
The highest-energy
processes which we witness in astrophysics provide us
an opportunity to test canonical noncommutativity in a way
that would be protected from the (however unlikely)
possibility of such a ``conspiracy" (UV structures that conspire
to cancel out the characteristic effects of canonical noncommutativity
in the infrared).

Of course one of the unknowns of our study is
the ``infrared scale" $1/(\theta \Lambda)$.
As a result,
the analysis of relevant astrophysics contexts at relatively
high energies will lead us to conclusions of the "either/or" type.
If a certain deformation of the dispersion relation is not seen
in observations conducted at an energy/momentum scale $E$
this can be interpreted in two ways:
one should either constrain the noncommutativity parameters accordingly
(to suppress the unwanted effect) or assume that
even at the relatively high energy scale $E$ some unknown
properties of the UV sector have managed to conspire to cancel out
the unwanted dispersive effects ({\it i.e.}
even the energy scale $E$ is below  $1/(\theta \Lambda)$).

In some of the astrophysical processes that we consider, in
addition to the high-energy particle there is also a very soft
particle, typically a soft photon. We will assume that these very
soft particles are unaffected by noncommutativity, since this is
consitent with the conservative approach we are adopting: we want
to test the idea of canonical noncommutativity relying exclusively
on its characteristic departures from commutative-spacetime predictions
at energy scales larger than $1/(\theta \Lambda)$.

For our programme of experimental searches of the effects
predicted by field theories in canonical noncommutative spacetime
another key issue is the one of the identification of the
SUSY-restoration scale ${\cal M}_{susy}$. While the phenomenology
we discuss is of even wider applicability, our key focus will be
on scenarios in which SUSY is present in the UV sector (in order
to contribute to the ``cure" of the infrared problems), but the
restoration of SUSY occurs at ultrahigh energies, possibly above
the noncommutativity scale $1/(\sqrt{\theta})$. In such a
situation SUSY can still effectively contribute to
the ``cure" of the infrared problems
of canonical noncommutativity, although it would occur at energy
scales typically higher than the ones usually considered for SUSY
restoration in commutative spacetimes affected by the ``hierarchy
problem". We feel that the role of SUSY in commutative spacetime
and in canonical noncommutative spacetime should be distinguished
more carefully than usually done in the literature. In
commutative spacetime Wilson decoupling between the UV and the IR
sectors is at work, and SUSY is only needed in order to provide
an elegant explanation for the hierarchy of mass scales present
in the Standard Model of particle physics, which would otherwise
be puzzling in light of the corresponding renormalization-group
results. In canonical noncommutative spacetime the
renormalization group analysis is alarming already before
considering problems
of mass-scale hierarchy: much more troubling problems are present
as a result of the lack of Wilson IR/UV decoupling. SUSY is
{\underline{desperately needed}} in order to cure some
{\underline{unacceptable}} IR problems, whereas in commutative
spacetime SUSY {\underline{improves the compellingness}} of our
description of particle physics, by eliminating the
(phenomenologically {\underline{acceptable}} but ``unnatural")
fine-tuning otherwise required by the mass-scale hierarchy.
In the study of
theories in commutative spacetime, because of the reasons just
mentioned, we have become accustomed to assuming a
SUSY-restoration scale of the order of $1TeV$ or $10 TeV$.
Theories in commutative spacetime are consistent even without
SUSY restoration, but assuming SUSY restoration at $1TeV$
or $10 TeV$ a more compelling picture emerges. In the new subject of
theories in canonical noncommutative spacetime we probably must
assume SUSY restoration, at least in the UV sector. Whether or
not theories in canonical noncommutative spacetime become more
compelling with a lower scale of SUSY restoration remains to be
established (it requires more studies in which the
renormalization group is applied to ``realistic" models based on
canonical noncommutativity).

\section{High-energy astrophysics observations and canonical noncommutativity}
As announced in Section~I, we are proposing to test the
predictions of canonical noncommutative spacetime using the same
techniques previously developed to search for the effects
possibly induced by the parameter $\lambda$ of the Lie-algebra
noncommutative spacetime (\ref{kmindef}). In fact, both classes of
noncommutative spacetimes are characterized by deformations of
the special-relativistic dispersion relation, and can therefore
both be tested using experiments with good sensitivity to such
deformations. We focus here on the role of deformed dispersion
relations for the evaluation of the threshold condition for
particle production in certain collision processes, and for the
dispersion of signals observed from distant astrophysical sources.
Since our objective here is the one of outlining a relatively wide
phenomenological programme, postponing to future studies a
detailed analysis of each of the proposals, we just list a few
experimental tests and provide estimates of the
sensitivity levels that appear to be within reach of these tests.
Our sensitivity estimates are also rather tentative: the relevant
astrophysics studies are experiencing a fast rate of improvement
and one can expect these sensitivities to increase significantly
over the next few years.

\subsection{Gamma-ray astrophysics}
It is a rather general feature~\cite{polonpap1} of ``quantized"
(discretized, noncommutative...) spacetimes to induce anomalous
particle-propagation properties. In some
pictures~\cite{grbgac,garaytest,gampul} of ``spacetime foam" (not
involving noncommutative geometry) one describes foam as a sort of
spacetime medium that, like other media, induces dispersion. As
discussed in the preceding Sections, in quantum pictures of
spacetime based on canonical or Lie-algebra noncommutative
geometry one also automatically finds deformations of the
dispersion relation. The observation that gamma-ray astrophysics
could be used to search for the effects of such deformed
dispersion relations was put forward in Ref.~\cite{grbgac},
focusing on foam-induced dispersion. The use of the same gamma-ray
astrophysics for tests of the predictions of $\kappa$-Minkowski
Lie-algebra noncommutative spacetimes was then discussed in
Refs.~\cite{polonpap1,gacqm100,dsr1dsr3}. Here we observe that
gamma-ray astrophysics can also be used to set bounds on the
$\theta_{\mu,\nu}$ parameters of canonical noncommutative
spacetimes. Bounds on deformations of the dispersion relation can
be set by analyzing time-of-arrival versus energy correlations of
gamma-ray bursts that reach our detectors coming from far away
galaxies. In presence of a deformed dispersion relation (which
implies that photons of different energies travel at different
speeds) photons emitted in a relatively short time (a burst)
should reach our detectors with a larger time spread, and the
spreading should depend on energy difference.

Gamma-ray bursts~\cite{piranGRBs} travel over
large distances, $\sim 10^{10}$ light years, and are observed
to maintain time-of-arrival correlation over very short
time scales, $\sim 10^{-3}$s.
For photons in the bursts that have energies in
the $100$KeV-$1$MeV range, as the ones observed by the BATSE
detector~\cite{piranGRBs}, this has of course allowed to
establish~\cite{grbgac,prlGRBschaef} that {\it in vacuo}
dispersion (if at all present) is small enough to induce relative
time-of-arrival delays between photons with energy differences of
a few hundred KeV that are below the $10^{-3}$s level. This turns
out to set a stringent limit on the $\lambda$ parameter of the
$\kappa$-Minkowski Lie-algebra spacetime, $\lambda < 10^{-30} cm$,
and experiments now in preparation will allow to probe even
smaller, subPlanckian, deformation
scales~\cite{grbgac,prlGRBschaef,glast}. Limits of comparable
significance can be obtained analyzing the bursts of photons
emitted by blazars~\cite{billetal}. An analysis of data on the
Markarian 421 blazar allowed to establish~\cite{billetal} that
photons with energies of a few TeV (and comparable energy
differences within the burst) acquire relative time-of-arrival
delays that are below the $10^3$s level for travel over distances
of order $100$Mpc. This again sets a limit on $\lambda$ that is
of order $\lambda < 10^{-30} cm$.

With respect to the analysis of these astrophysics contexts
in $\kappa$-Minkowski Lie-algebra spacetime, the analysis
in canonical
noncommutative spacetimes is complicated by the polarization
dependence of the deformation of the dispersion relation
for photons.
However, it appears safe to assume that the emissions by
gamma-ray-bursts and blazars are largely unpolarized, and
therefore canonical noncommutative spacetimes would imply that at
least a portion of the bursts (also depending on the position of
the source, which of course fixes the direction of propagation of
the photons that reach us from the source) would manifest a
dispersion-induced effect.
A short-duration light burst
travelling in a birefringent medium increases its time spread
over time.

As announced we just
want to estimate this type of sensitivities and we want to focus on data
that are obtained as far from the infrared as possible. Considering the TeV
photons of the mentioned blazar and
the nominal~\cite{billetal} $\delta T/T \sim 10^{-12}$ accuracy
(with $\delta T$ the observed level of time-of-arrival simultaneity
and $T$ the overall time of flight),
the $(p \theta)^{-2}$ behaviour of the
deformation of the dispersion relation\footnote{In our estimates
we take the numerical factors $\zeta$ introduced in Section~II to be of
order 1. They are probably smaller than 1 (they involve the QED coupling
constant $\alpha$) but at this preliminary stage we are only
trying to establish a rough picture of the $\theta$-sensitivities
of some relevant experiments, and we would not be too concerned
even with inaccuracies of 1 or 2 orders of magnitude.}
leads to expected (polarization-depedent and) energy-dependent
time delays of $(p^2 \theta)^{-2} T \sim (TeV^2 \theta)^{-2} T$.
Therefore, if terms of the type $(p \theta)^{-2}$ are present in the
dispersion relation that applies to photons at energies in the TeV range,
then necessarily one must impose $\sqrt{\theta} > 10^{-14}cm$,
{\it i.e.}  $1/\sqrt{\theta} < 1 GeV$
(this assures that $(TeV^2 \theta)^{-2} \ll 10^{-12}$,
so that the observed level of time-of-arrival simultaneity is not
in conflict with canonical noncommutativity).

One can confidently exclude\footnote{We are making the (apparently robust)
assumption that is not possible to make a working phenomenological
proposal based on a noncommutativity
energy scale $1/\sqrt{\theta}$ with value that falls within the energy
scales to which we do have access in laboratory experiments,
such a particle accellerators.}
spacetime noncommutativity at the $GeV$
scale, and therefore this analysis imposes that even for $TeV$ photons
the $(p \theta)^{-2}$ behaviour of the
deformation of the dispersion relation
is not acceptable phenomenologically.
One must therefore either reject canonical noncommutativity
alltogether or assume that the UV sector is effective in eliminating
the $(p \theta)^{-2}$ corrections all the way up to the $TeV$ scale
(which can be accomplished by introducing suitable SUSY in the UV).

In a certain sense our result is negative, since it
does not leave us with any hope of finding one day the effects
of canonical noncommutativity through this gamma-ray-dispersion studies.
A positive result would have
been to find present limits on TeV gamma-rays dispersion
to be consistent with some still relatively high noncommutativity
energy scale $1/\sqrt{\theta}$. This would have left us with
the hope that future more sensitive searches of TeV gamma-rays dispersion
might find some evidence of dispersion and that this evidence could
be interpreted in terms of canonical noncommutativity.

\subsection{Hard-photon FIRB-absorption threshold}
Both in the study of spacetime-foam
models~\cite{grbgac,garaytest,gampul} and in the study of
$\kappa$-Minkowski Lie-algebra noncommutative spacetimes the
emerging deformed dispersion relations have also been
analyzed~\cite{gacqm100,dsr1dsr3,gactp,kifu,aus,kluz}
for what concerns their implications for the
determination of the threshold conditions for particle-production
in collision processes.
The interested reader can find these previous results
in Refs.~\cite{gacqm100,dsr1dsr3,gactp,kifu,aus,kluz},
and references therein.
Here we apply the same line of analysis to
the deformed dispersion relations that emerge in canonical
noncommutative spacetimes. An interesting process for this type
of threshold analyses is electron-positron pair production in
photon-photon
collisions: $\gamma + \gamma \rightarrow e^- + e^-$.
This is a significant process for the physics of blazars.
Certain blazars, and in particular the Markarian 501 blazar,
are known to emit photons up to very high
energies. For photons with energies of a few $TeV$ the soft
photons of the Far Infrared Background Radiation (FIBR) provide
viable targets for electron-positron pair production. The hard
photons could therefore disappear in an electron-positron pair.

This hard-photon absorption phenomenon is analyzed using a simple
kinematical requirement. The balance of the hard photon energy $E$
and the soft FIBR photon energy $\epsilon$
should satisfy the kinematical condition
for the production of an electron-positron pair: $E > m_e^2 / \epsilon$.
The end result is that (with some uncertainty due to our partial
knowledge of the FIRB) absorption by the FIBR should become
significant for hard photons of energies of $10 TeV$ and higher.

Observational evidence with respect to this absorption prediction
is still preliminary
but a cut off on the hard-photon observations from blasars,
which would signal absorption, is being reported~\cite{krennMKcut,ahaMKcut}
at about the right energy scale. While this evidence is probably
not yet conclusive~\cite{aus,gactp,steckernoparadox,berezin},
we adopt here a conservative perspective
and take as working assumption that the expected cutoff is being
seen, as expected, and therefore any deformation of the dispersion relation
should be small enough, at energies in the 10 TeV range, that the
standard absoprtion prediction is not significantly affected.

In the case of canonical noncommutative spacetimes it
is important to take into account the polarization dependence of
the deformation of the photon dispersion relation. This would
lead to different threshold conditions for different
polarizations. If the threshold is increased only for photons
with a certain polarization, then one would expect that at the
conventional special-relativistic threshold there would be a
suppression of the flux, but only partial. Another delicate issue
is the one of the very soft (FIBR) photon involved in the
process. As discussed in the previous sections, we will neglect
possible\footnote{In principle, canonical noncommutativity could
lead to large deformations for infrared photons, but the
corresponding predictions are subject to severe modification
by the UV sector, through IR/UV mixing. Moreover, at infrared
energies we have a huge amount of data confirming the
special-relativistic behaviour of photons, and therefore (if
canonical noncommutativity must at all be studied) an {\it a
priori} assumption of phenomenology in canonical noncommutative
spacetimes must be that the unknown UV sector should cure the
potential problems of the far infrared.}
canonical-noncommutativity deformations of the dispersion
relation for such low-energy photons.

We shall therefore assume that only the hard
photon is affected by the deformed dispersion relation, while the FIBR
photon involved in the process obeys the ordinary special-relativistic
dispersion relation. Again, postponing a more detailed analysis to future
studies, we estimate here the range of the $\theta_{\mu,\nu}$ parameters
that could significantly affect the absorption threshold
for multi-TeV photons from blazars.
Since the deformation of the dispersion relation is governed by correction
terms of the type $(p \theta)^{-2}$, it is easy to verify that the
absorption threshold could be significantly affected only
if $(p \theta)^{-2} \gtrsim m_e^2$. In the case of
interest $p \sim 10$TeV, and therefore the threshold can be
significantly affected only if the
value of $\sqrt{\theta}$ is
at least as large as $10^{-15}cm \sim (10^{-2} TeV)^{-1}$.
Again we are finding a limit on the noncommutativity
energy scale, $1/\sqrt{\theta} < 10 GeV$,
which on other grounds (see previous Subsection)
one can confidently exclude.
Therefore, if indeed more refined data confirm
that the threshold for multi-TeV photons from blazars is correctly
predicted by the ordinary special-relativistic dispersion relation,
this analysis imposes that even for photons of energies of
order $10 TeV$ the $(p \theta)^{-2}$ behaviour of the
deformation of the dispersion relation
is not acceptable phenomenologically.
One must therefore either reject canonical noncommutativity
alltogether or assume that the UV sector is effective in eliminating
the $(p \theta)^{-2}$ corrections all the way up to the $10 TeV$ scale
(which can be accomplished by introducing suitable SUSY in the UV).

\subsection{Cosmic-ray threshold}
Ultra-high-energy cosmic rays can interact with
the Cosmic Microwave Background Radiation (CMBR),
producing pions ($p + \gamma \rightarrow p + \pi$).
This phenomenon can be analyzed
just like the pair-production process relevant for observations
of multi-$TeV$ photons, which we discussed in the previous
Subsection. Taking into account the typical energy of CMBR
photons, and assuming the validity of the kinematic rules for the
production of particles in our present, classical and continuous,
description of spacetime (ordinary relativistic kinematics
and dispersion relation),
one finds that these interactions should lead to an upper limit
$E < 5 {\cdot} 10^{19}$eV, the GZK limit~\cite{gzk}, on the
energy of observed cosmic rays. Essentially, cosmic rays emitted
with energies in excess of the GZK limit should lose energy on the
way to Earth by producing pions, and, as a result, should still
satisfy the GZK limit when detected by our observatories.

As for the case of the multi-$TeV$ photons,
a deformed dispersion relation can affect
the prediction
for the $p + \gamma \rightarrow p + \pi$ threshold.
Again we refer the reader interested in the analysis of the
cosmic-ray paradox within other quantum-spacetime frameworks to
previous results in the
literature~\cite{gacqm100,dsr1dsr3,kifu,gactp,emnthresh,barcelona};
here we just intend to focus on canonical noncommutative
spacetimes and estimate values of the $\theta_{\mu,\nu}$ parameters
that could significantly affect the GZK cosmic-ray threshold.

The process $p + \gamma \rightarrow p + \pi$ does not involve
any hard photons, and, for the reasons discussed above, we will
assume that the soft CMBR photon should be analyzed according to
the conventional special-relativistic dispersion relation.
For the proton, which is electrically charged, strong anomalies
in the dispersion relation are not expected.
The most significant dispersion-relation deformation
relevant for this process could be attributed to the neutral/uncharged
pion\footnote{Our wording here is rather prudent as a result
of the fact that the pion is a composite particle. In Lie-algebra
noncommutative spacetimes it appears\cite{dsr1dsr3}
that composite particles are subject to a deformed
dispersion relation that is different from the one for fundamental
particles. We are not aware of analogous results in canonical
noncommutative spacetimes, but of course this issue would be
important for the kinematics of $p + \gamma \rightarrow p + \pi$.}.
Assuming that $(p \theta)^{-2}$ corrections to dispersion relations
do indeed characterize the kinematics of this process, we observe
that these deformations would be significant
at the GZK scale if $(p \theta)^{-2} \gtrsim m_p m_\pi$,
with $p \sim 10^{20}$eV.
Therefore, for $\sqrt{\theta} < 10^{-20}cm \sim (10^2 TeV)^{-1}$
there could be significant implications for this type
of experimental studies.

This possible implication of canonical noncommutativity
certainly deserves interest.
In this cosmic-ray context the prediction of the standard
special-relativistic dispersion relation can be questioned.
As mentioned this standard dispersion relation leads to the
GZK limit, and instead, observations of several cosmic-rays
above the GZK limit (with energies as high as $3 {\cdot} 10^{20}$eV)
have been reported~\cite{AgaWat}.
The fact that canonical noncommutativity at or above
the  $1/\sqrt{\theta} \sim (10^2 TeV)$ scale
could significantly affect the GZK prediction could be used in
attempts to explain these surprising violations
of the standard cosmic-ray paradox.
Canonical noncommutativity with $1/\sqrt{\theta} \sim (10^2 TeV)$
also appears to be plausible, since we do not yet have access
to the $10^2 TeV$ scale in laboratory experiments.

In order for the dispersion relation for particle
of energies $\sim 10^{20}$eV
to be affected by $(p \theta)^{-2}$ corrections
it is necessary~\cite{gianlucaken} to assume
that SUSY is not yet restored at $10^{20}eV = 10^{8}TeV$.
This is consistent with the experimental information so far
available, but it would be in conflict with the expectations
thaat are formulated in the most popular theoretical scenarios
for particle physics in commutative spacetime.
As emphasized in Section~III,
the role of SUSY in commutative spacetime
and in canonical noncommutative spacetime should be distinguished
more carefully than usually done in the literature. In
commutative spacetime Wilson decoupling between the UV and the IR
sectors is at work, and this has a nontrivial role in the
expectation that SUSY should be restored well below $10^{8}TeV$
(probably already at a scale of a few $TeV$s).
In canonical noncommutativity there is no Wilson IR/UV decoupling
and it appears plausible to contemplate a much higher
SUSY-restoration scale, possibly above $10^{8}TeV$.

\subsection{1987a-type supernovae}
Our final observation concerning possible
experimental investigations of the dispersion-relation
deformations that emerge in canonical noncommutative spacetimes
involves neutrinos. We have seen in Section~II that also neutral
spin-1/2 particles could acquire a dispersion-relation
deformation, but with a somewhat softer (logarithmic, rather than
inverse-square power) dependence on ``$p \theta$".
This might mean that, if any of these experimental searches
ends up being successful, the first positive results are unlikely
to come from data on neutrino kinematics. Still,
one should keep in mind that neutrinos are a particularly
clean spacetime probe. The identification of
quantum-spacetime effects that can
be qualitatively described as ``spacetime-medium effects",
such as the ones induced by canonical noncommutativity\footnote{As mentioned,
canonical noncommutative spacetimes require the existence of
a preferred frame, and most of their features, especially concerning
the associated dispersion-relation deformations, can be described
in close analogy with ordinary (commutative-spacetime) physics
in presence of a medium (think, for example, of the laws that
govern propagation of light in certain crystals).},
can sometime be obstructed by the fact that most particles
also interact with more conventional (electro-magnetic) media,
as indeed is the case for the particles that reach our detectors
from far-away galaxies.

Neutrinos are only endowed with weak-interaction charges.
They are therefore mostly insensitive
to this conventional dispersion-inducing effects, and as a result
could provide clean signatures of spacetime-induced
dispersion.\footnote{Interestingly, if spacetime was indeed described
by a canonical noncommutative geometry, neutrinos would acquire at once
both a deformed dispersion relation and the ability to
interact, however softly, with electromagnetic media.}
In this respect supernovae of the type of 1987a
might provide a interesting laboratory because of
the relatively high energy of the observed neutrinos ($\sim 100$MeV),
the relatively large distances travelled ($\sim 10^3,~10^4$light years),
and the short (below-second) duration of the bursts.

\section{Closing remarks}
We have argued that certain types of experimental tests which
were previously considered in the literature on Lie-algebra
noncommutative spacetimes, can also be of interest
for investigations
of canonical noncommutative spacetimes.
The relevance of these experimental contexts
comes from the fact that
they rely on deformed dispersion relations,
a feature that is present in both types of noncommutative spacetimes.

The observations involve particles of relatively high energies.
This is welcome
in Lie-algebra noncommutative spacetimes because the effects are fully
confined to the high-energy regime, and we argued that it should also
be welcome in canonical noncommutative spacetimes in light of the
peculiar dependence of the infrared sector on the unknown structure
of the UV sector. With low-energy tests of canonical noncommutativity
we probe predictions of the theory which are highly sensitive to
the structure of the unknown UV sector. If one manges to gain access
to observations that pertain the regime $E > 1/(\Lambda \theta)$
the unwanted sensitivity to the UV sector would be avoided.

In two types of observations involving TeV and multi-TeV photons
we were led to the conclusion that
(if not rejected alltogether)
requires a UV sector which (through IR/UV mixing) eliminates
the $(p \theta)^{-2}$ corrections to the dispersion relation
all the way up to the $10 TeV$ scale
(which can be accomplished by introducing suitable SUSY in the UV).

Potentially more exciting is our preliminary analysis of
ultra-high-energy cosmic rays from the canonical-noncommutativity
perspective. We found that the type of deformed
dispersion relations predicted by canonical
noncommutativity could explain the puzzling observations of
cosmic rays above the GZK cutoff ($5 {\cdot} 10^{7} TeV$).
This would require a noncommutativity scale in the
neighborhood of $1/\sqrt{\theta} \sim 10^2 TeV$.
We clarified that
in order to obtain a working canonical noncommutativity
scenario with this attractive prediction
it appears necessary to obtain a satisfactory description
of pions as composed by fundamental quarks (a description
which is turning out to troublesome in other noncommutative
spacetimes) and a satisfactory canonical-noncommutativity
physical picture in which
SUSY is restored only at scales higher than $10^{8} TeV$.

\bigskip
\bigskip
\bigskip

\section*{Acknowledgements}
The present manuscript is an updated version of
http://arXiv.org/abs/hep-th/0109191v1.
With respect to the experimental information available
at the time of writing http://arXiv.org/abs/hep-th/0109191v1
the situation has evolved
significantly. The present updated description of the experimental
situation also benefitted from conversations with Paolo Lipari.
Our updated remarks on the IR/UV mixing also benefitted from the
results of Ref.~\cite{gianlucaken}, for which GAC worked in
collaboration with Gianluca Mandanici and Kensuke Yoshida.
GAC also thankfully acknowledges conversations with Roman Jackiw,
Alan Kostelecky and Lorenzo Marrucci.
SN acknowledges KOSEF 2000-1-11200-001-3 and also the
BK21 program of the Korea Research Fund (2001).

\end{document}